\documentclass[10pt, journal]{IEEEtran}
\usepackage{comment}
\usepackage{multirow}
\usepackage{color}
\usepackage{wrapfig}
\usepackage{pifont}

\usepackage{subfigure}
\usepackage{enumitem}
\usepackage{url}
\usepackage{cite}
\usepackage{bm}
\usepackage{graphicx}
\usepackage{amsmath}
\usepackage{amssymb}
\usepackage{newfloat}
\DeclareFloatingEnvironment{Definition}
 \usepackage{booktabs}
\definecolor{darkgreen}{rgb}{0.0, 0.5, 0.0}
\definecolor{darkblue}{rgb}{0, 0, 0.543}
\definecolor{darkcyan}{rgb}{0, .543, .543}

\color[rgb]{1.0, 1.0, 1.0} 
\usepackage[compact]{titlesec}
    \titlespacing{\section}{0pt}{2ex}{1ex}
    \titlespacing{\subsection}{0pt}{1ex}{1ex}
    \titlespacing{\subsubsection}{0pt}{1ex}{1ex}
\usepackage[table]{xcolor}

\allowdisplaybreaks
\title{Non-Intrusive Driver Behavior Characterization From Road-Side Cameras}

\begin{document}

\bstctlcite{IEEEexample:BSTcontrol}

\author{
Pavana Pradeep Kumar, Krishna Kant and Amitangshu Pal
\thanks{Pavana Pradeep Kumar and Krishna Kant are with the Department of Computer and Information Sciences, Temple University, Philadelphia, PA, USA. (e-mail:  pavana.pradeep@temple.edu, kkant@temple.edu)}
\thanks{Amitangshu Pal is with the Computer Science and Engineering, Indian Institute of Technology Kanpur, Kanpur, India. (e-mail:  amitangshu@cse.iitk.ac.in)}
}
\maketitle

\begin{abstract}
In this paper, we demonstrate a proof of concept for characterizing vehicular behavior using only the roadside cameras of the ITS system. The essential advantage of this method is that it can be implemented in the roadside infrastructure transparently and inexpensively and can have a global view of each vehicle's behavior without any involvement of or awareness by the individual vehicles or drivers. By using a setup that includes programmatically controlled robot cars (to simulate different types of vehicular behaviors) and an external video camera set up to capture and analyze the vehicular behavior, we show that the driver classification based on the external video analytics yields accuracies that are within 1-2\% of the accuracies of direct vehicle-based characterization. We also show that the residual errors primarily relate to gaps in correct object identification and tracking and thus can be further reduced with a more sophisticated setup. The characterization can be used to enhance both the safety and performance of the traffic flow, particularly in the mixed manual and automated vehicle scenarios that are expected to be common soon. 

\end{abstract}


\section{Introduction}

Road accidents are responsible for $\sim$5 million severe injuries and $\sim$50K deaths annually in the USA (https://www.bankrate.com/insurance/car/car-crash-statistics/).  According to the National Highway Traffic Safety Administration (NHTSA), at least one of the following risky behaviors was evident in 45\% of fatal crashes involving passenger vehicles: speeding, alcohol impairment, or not keeping a safe distance, etc~\cite{stewart2022overview}.  Intelligent Transportation Systems (ITS) aim to reduce accidents and make traffic flow smoother by introducing capabilities to monitor traffic and communicate with vehicles. 

Characterizing driver behavior is essential for improving traffic safety and performance.  It should be noted that the term ``driver" does not necessarily imply a manually operated vehicle.  As the market for automated vehicles develops and becomes significant, it is anticipated that vehicles from different manufacturers will have distinct personalities, influenced in part by the design of the software and in part by the ``personalization" that users will 
desire~\cite{bruck2021investigation}. Some users may be comfortable with automated driving aggressively, while others prefer a more conservative driving style.  This aspect also depends on familiarity with automated driving and driving conditions.  Therefore, in this paper, driver behavior refers to vehicular behavior, although the type of feedback provided will still be different for manual vs. automated vehicles.

\subsection{Vehicular and Driver Behavior Characterization}

There is a tremendous amount of work on vehicular behavior characterization, mainly regarding monitoring vehicular kinematics parameters.  Vehicular behavior can be measured either from the sensors in the car through the OBD (onboard diagnosis) port or a device carried/mounted in the car.  Typical behavioral measurements include acceleration, braking, distance to the next car, speed and speed variability, lane change behavior, etc~\cite{martinez2017driving}. 

In addition, there is a large body of work specifically targeting the behavior of the human driver in terms of direct actions (use of brake, accelerator, or steering wheel), alertness, and physical health (e.g., eye or head movements, facial expressions, lack of focus on the road, etc.) to determine if the driver is tired, drowsy, etc.~\cite{ramzan2019survey}. 
Other sensors may also be used for such things as analyzing breathing patterns and breath smell (for alcohol), and may even use contact sensors such as those for measuring heart rate, brain signals, skin conductance, etc.  With their increasing array of sensors and wearable devices such as smartwatches, smartphones can easily measure many vehicular and driver parameters.  The measurements are generally used to train a classifier to categorize the driver.

However, such methods have several inherent problems.  First, the measurements require either the deployment of special gadgets in the vehicles or depend on the driver carrying or wearing devices with appropriate sensors.  Using devices carried by a driver for a driver or vehicular behavior is impractical and has little value beyond a proof of concept.  While the vehicles do monitor all of the kinematic parameters and thus could estimate the driver behavior, this has limitations since each vehicle only has a local view.  The behavior of a group of vehicles traveling together or in close proximity is interdependent, and these interdependencies make it challenging to identify the cause-and-effect relationships among the vehicles.  For example, a sudden slowing of a vehicle will cause the vehicle behind it to also slow suddenly.  While the sudden slowing of the front vehicle may be undesired (e.g., aggressive driving), that of the vehicle behind is highly desired for maintaining safety.  Another serious problem with these methods is the awareness of the drivers that their driving style or physical condition (e.g., drowsiness, drunkenness) is being monitored.  This intrusiveness would surely change driver behavior; thus, the monitoring will be inaccurate.  Furthermore, many drivers may refuse such monitoring or try to defeat it.

\subsection{Our contributions:}

In this paper, we demonstrate that it is possible to characterize vehicular behavior very accurately using only the roadside cameras of the ITS system.  To the best of our knowledge, this is the first such study; its key advantage is that it is entirely non-intrusive, inexpensive, and does not impact driving behavior in any way.  Also, the global view can deduce the more obvious cause-and-effect relationships between adjacent vehicles (e.g., whether the car decelerated on its own or due to diminishing distance from the vehicle ahead).  Furthermore, the minimal error rate of our video analytics system is attributable to instances where the vehicular ID is not tracked correctly or the camera is too far from the vehicle.  These errors can be further minimized by more elaborate feature-based tracking of vehicles (e.g., by color, shape, etc.) and by having associations between multiple cameras.  The monitoring can be used to provide advice to the vehicles in terms of enhancing safety and traffic performance.  The scheme becomes even more effective with automated vehicles since they can help smooth out the traffic by behaving in a specific way. 

\subsection{Paper outline:}

The remainder of the paper is organized as follows. Section~\ref{s:related-work} discusses the related work. Section~\ref{s:overall-framework} presents our proposed framework for driver behavior characterization. Experimental evaluation and results are summarized in section \ref{s:experimental-evaluation}. Section \ref{s:conclusions} then concludes the paper.

\section{Related work}
\label{s:related-work}

Different methods that have been studied for identifying driver behavior can be classified into broadly two categories: using (a) in-vehicle sensor or video data analysis and (b) analyzing driver's physiological data. In the following, we summarize them separately.

\textbf{Sensor or video data analysis:} Authors in~\cite{5482295} have developed a drunk driving and alerting system using a mobile phone that can be installed inside a car; the phone can record the acceleration samples from the phone sensors and compare them with the drunk driving patterns. In~\cite{6083078}, the authors have proposed an aggressive driving behavior detection system that uses data from multiple smartphone-based sensors (i.e., accelerometer, gyroscope, magnetometer, GPS, video) and recognizes abnormal driving behaviors using the Dynamic Time Warping (DTW) algorithm. Similarly, the authors in~\cite{app112110420} have developed a driver behavior classification and sharing system based on vehicle-mounted acceleration samples. Authors in~\cite{app112210562} have used a public dataset named UAH-DriveSet~\cite{7795584} to analyze different in-vehicle sensory data and classify driver behaviors using different machine learning based classification methods.
In~\cite{4579980}, the authors have studied driver fatigue detection using eye tracking, where distances between the intensity changes in the eye area are measured to determine whether the eyes are closed or not. 
Similar other studies~\cite{5337166,ITO200120} have analyzed various in-vehicle video data like driver's facial expressions, eye or head movements, etc., for determining driver's behaviors.

\textbf{Physiological data analysis:} Psycho-physical states of drivers using respiration rates and ECG signals can also be studied through wearable devices, which also determine drivers' stress or distraction levels. 
In~\cite{lal_craig_2002}, the authors have shown that significant electroencephalographic and psychological changes (like increased delta and theta activity, lower heart rate, changes in blink rate, etc.) occur during fatigue.
In~\cite{6025569}, the authors have used psychological features, such as EEG and ECG signals to classify the driver's behavior into four categories (alert, mild fatigue, deep fatigue, and drowsiness) using a support vector machine (SVM). Similar studies on physiological data analysis for driver's state detection are reported in~\cite{10.1007/978-3-319-19258-1_8}.

The existing works for driver behavior detection require either installation of in-vehicle sensors and cameras or require on-body wearable sensors, which bring additional installation overhead. As opposed to the existing literature, our proposed framework characterizes driver behavior from roadside camera frames, making the solution inexpensive, non-intrusive, and easily deployable.

\begin{figure*}
\center
\includegraphics[width=0.8\linewidth]{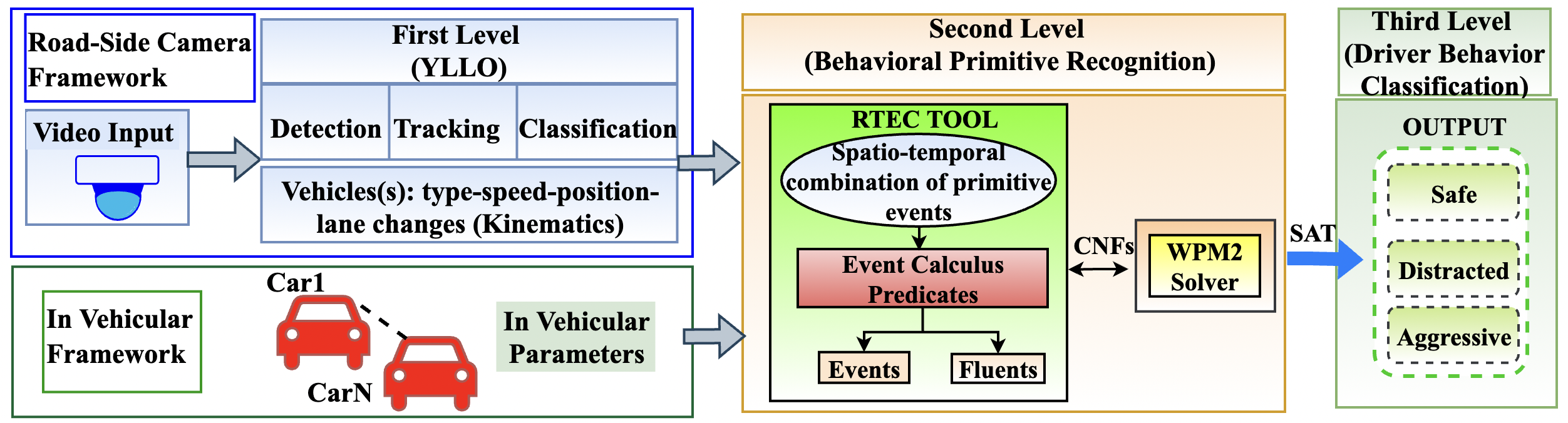}
\caption{Driver Behavior Characterization Framework.}
\label{f:architecture}
\vspace{-12pt}
\end{figure*}

\section{Framework for Driver Behavior Prediction}
\label{s:overall-framework}

\subsection{Categorizing Driver behavior}
\label{s:catergories-driver}

Driver behavior is important from the safety perspective, but there is no standard way to characterize drivers. In this paper, we adopt a 3-way classification, where a driver is designated as safe, aggressive, and distracted~\cite{U.S.NHTSA}. Please note that our ``safe" category includes any behavior that is not considered aggressive or distracted; therefore, there is no undetermined behavior. For this categorization, we define a small set of ``micro-behaviors" and then characterize the driver based on the combination of micro-behaviors observed. The micro behaviors include: (a) \textit{Weaving}, or driving alternatively towards one side of the lane and then the other; (b) \textit{Sudden Steer} where the driver makes an abrupt redirection when driving along a straight course; (c) \textit{Hard Braking}; (d) \textit{Lane Drifting}, or not keeping in the center of the lane while driving; (e) \textit{Straddling}, or driving while staying close to one side of the lane; and (f) \textit{Over speeding}, or driving significantly above the speed limit. By performing object detection and estimating driving parameters over traffic videos as well as in vehicular data, we observe these micro-behaviors and classify the driver's behavior into one of three categories. For example, an aggressive driver can be identified by a combination of the following behaviors: weaving through traffic, driving at a higher speed (speeding), following other vehicles too closely with rapid steering, and applying harsh braking.

In this paper, we formulate the driver behavior detection problem as a boolean satisfiability problem so that the popular SMT (Satisfiability Modulo Theory) based tools can be used along with suitable theories. We make use of a popular framework called {\em Event Calculus} (EC) that introduces the concept of Events, which are actions that occur at a specific point in time, and Fluents, which are entities whose state changes in response to the occurrence of an event or action. EC provides constructs to reason about situations, events, and changes in time, allowing a precise specification of time relationships between situations and events, which is essential for describing the various driving style patterns exhibited by drivers. In addition, concurrent actions can also be specified in EC.

\subsection{System Architecture}

We assume that the camera deployment is such that all vehicles on the roadway are visible without excessive image distortion. Given the increasing processing power in smart cameras, each camera can do the object detection and tracking tasks discussed here rather than sending the raw video feed to the next level known as {\em Road-Side Edge Controller (REC)}. In our earlier work, we designed a lightweight object recognition and tracking algorithm called YLLO (You Look Less than Once) that can run in the cameras and avoid transmission of redundant frames to RECs~\cite{Pavana-YLLO}. The RECs receive video streams from multiple cameras along a road segment and use them to monitor driver behaviors and associated anomalies. Further processing, including perspective transformation and estimation of orientations and speeds of the objects, may be done by the camera itself or by the RECs. The REC can then build a spatio-temporal logic model of the situation that includes all ``facts'' of different driver behaviors and the conditions leading to near-miss accidents along with the supporting ``theories'' (i.e., Newton's laws, arithmetic, etc.)

Fig.~\ref{f:architecture} depicts our overall architecture, which is comprised of two different frameworks. One framework is based on Deep Learning (DL) and reasoning that runs on roadside cameras receiving traffic videos as input. The other is based on logical reasoning that runs on individual vehicles receiving in-vehicle parameters as inputs. 

The initial stage of the roadside camera framework is a lightweight object detection and tracking model based on a convolutional neural network (CNN) model called YLLO that we have developed for video analysis~\cite{Pavana-YLLO}. YLLO performs processing of video sequences in a highly efficient manner. The second stage for each detected object (e.g., vehicles, pedestrians, etc.) is a spatio-temporal logic-based reasoning system that captures the relative movements of the objects in real-time to identify various driver behaviors. The reasoning framework operating on individual vehicles consists of a formal specification of driver behavior used by an event recognition tool.

\subsection{YLLO based Object Detection and Tracking}
\label{s:yllo}

YLLO is a  lightweight object detection technique based on YOLOv6 and is optimized for continuous video streams by utilizing redundancy to identify the ``only" essential frames. The previous version of YLLO as in~\cite{Pavana-YLLO} was based on YOLOv4; for this work, YLLO has been updated to a newer and more efficient version among all versions of YOLO, YOLOv6~\cite{li2022yolov6}. YLLO is a three-stage process that begins with a scene change detection algorithm and progresses to object detection via YOLOv6. The Simple Online and Real-time Tracking with a Deep Association Metric (Deep-SORT) algorithm assign a tracker to each detected object or multiple objects. YLLO decouples classification and regression tasks to eliminate redundant objects between the frames. Additionally, before sending frames to object detection, for the scene change detection, it generates Color Difference Histograms (CDH) for edge orientations, where edge orientations are determined using the Laplacian-Gaussian edge detection framework.

\subsection{Spatio-Temporal Reasoning}

In our work, we use an efficient dialect of the Event Calculus, termed ``Event Calculus for run-time Reasoning" (RTEC)\cite{artikis2014event}. RTEC is an open-source implementation of the EC in Prolog and uses LTL (linear time logic) with integer time points. RTEC implements novel techniques for identifying complex events from a set of micro-behaviors and is scalable to large volumes of complex events.  

We can define micro-behavior in RTEC as rules that define the event instances using the predicates ``happens at" (hA) and ``happens for" (hF). The {\em fluents} that are time-varying properties and the effects of events on {\em fluents} are defined using the \textit{inA} and \textit{tA} predicates. The value of {\em fluents} at any time point is defined using the \textit{hoA} and \textit{hoF} predicates. If {\em F} is a variable ranging over {\em fluents}, the term {\em F=V} denotes that variable {\em F} has a value {\em V}. There also exists Boolean fluents with values \textit{true} or \textit{false}. Table \ref{tab:RTEC-predicates} shows the predicates used in the RTEC tool.

\begin{table}[htb]
\vspace{-12pt}
\footnotesize\centering
\caption{Table: Main predicates of RTEC}
\label{tab:RTEC-predicates}
\begin{tabular}{|l|p{2.5in}|}
\hline
\multicolumn{1}{|c|}{Predicate} & \multicolumn{1}{c|}{Meaning} \\ \hline
\textit{hA}(E, T)               & Event E {\em happens at} time T \\ \hline
\textit{hF}(E, I)               & Event E {\em happens for} I intervals \\ \hline
\textit{in}(F=V)             & {\em Initial} value of fluent F =V at time 0 \\ \hline
\textit{hoA}(F=V, T)            & Value of fluent F {\em holds at} V at time T \\ \hline
\textit{hoF}(F=V, I)            & Value V of fluent F {\em holds for} I intervals continuously \\ \hline 
\textit{inA}(F=V, T)        & Fluent F with value V {\em initiated at} time T
\\ \hline 
\textit{tA}(F=V, T)       & Fluent F with value V {\em terminated at} time T
\\ \hline
\end{tabular}
\end{table}

The characterization of real-world behavior often involves fuzziness, whereas a Boolean logic framework requires an assertion to be either true or false. It is not straightforward to extend RTEC to fuzzy or probabilistic logic; therefore, we model the fuzziness indirectly by introducing weights. Recall that we classify driver behavior in terms of micro-behaviors, some of which are essential for characteristics of certain driver behavior. For example, sudden braking can be considered a key characteristic of aggressive driving. Thus we consider certain micro-behavioral assertions as {\em hard} in that they must hold, whereas others can be considered as {\em soft} or optional. Let $S_k$ denote the set of soft assertions for driver behavior $k$. We define a weight, denoted as $w_{ik}$ for each member $i$ of the set $S_k$. Then, the driver behavior $k$ will be recognized as (a) all hard assertions (micro-behaviors) holding, and (b) $\sum_{i\in S_k}i w_{ik}>W_k$ for some threshold $W_k$. Note that the weights need not be static but depend on various spatio-temporal factors and context (e.g., day vs. night time, roads with different speed limits, etc.) A weight change would require pausing all current condition evaluations, changing all weights that need to be changed simultaneously, and then restarting the evaluation. To handle hard/soft conditions, we can use an extension to the Boolean Satisfiability problem known as {\em Weighted Partial Maxsat} (WPM2). In WPM2, each clause is designated as either {\em hard} or {\em soft} with a given weight. We then have an optimization problem to find an assignment that satisfies all hard clauses and minimizes the total weight of soft clauses. The SAT core returned by the WPM2 solver confirms the presence of driver behavior.

\subsection{Defining Events and Fluents in RTEC}

In second stage of the framework, the RTEC tool receives input as event calculus (EC) predicates representing time-stamped micro behaviors detected on individual video frames or the recorded in-vehicular parameters as shown in Fig~\ref{f:architecture}. For example, the object's bounding box coordinates can define the appearance of a static object or multiple moving objects in each frame. We also have the object's angle/orientation and the direction in which they are moving, as well as lane drifting, which indicates if the vehicle is not staying in the center of the lane.

\begin{table}[t]
\footnotesize\centering
\caption{Table: Events and Fluents Defined}
\label{tab:evt-fluents}
\begin{tabular}{|l|p{2.2in}|}
\hline
\multicolumn{2}{|l|}{\textbf{Events Defined}}                                                                      \\ \hline
\multicolumn{1}{|l|}{\textit{proximityLeft(v)}}     & vehicle \textit{v} is close to the left side of the lane              \\ \hline
\multicolumn{1}{|l|}{\textit{proximityRight(v)}}    & vehicle \textit{v} is close to the right side of the lane             \\ \hline
\multicolumn{1}{|l|}{\textit{laneDrifting(v)}}      & vehicle \textit{v} is not keeping the center of the lane              \\ \hline
\multicolumn{1}{|l|}{\textit{changeLane1(v)}}       & vehicle \textit{v} changes to the lane1 to due to a steering maneuver \\ \hline
\multicolumn{1}{|l|}{\textit{changeLane2(v)}}       & vehicle \textit{v} changes to the lane2 to due to a steering maneuver \\ \hline
\multicolumn{1}{|l|}{\textit{weaving(v)}}       & vechicle \textit{v} driving alternatively on one side of the lane and then the other            \\ \hline
\multicolumn{1}{|l|}{\textit{suddenSteer(v)}}       & vehicle \textit{v} applies a sudden steer                             \\ \hline
\multicolumn{1}{|l|}{\textit{straddling(v)}}    & vehicle \textit{v} is positioned over lane lines rather than between them                       \\ \hline
\multicolumn{1}{|l|}{\textbf{Fluents Defined}}      & \textbf{Meaning}                                             \\ \hline
\multicolumn{1}{|l|}{\textit{atLane1(v)}}           & the existence of vehicle \textit{v} in lane 1 in the scenario         \\ \hline
\multicolumn{1}{|l|}{\textit{atLane2(v)}}           & the existence of vehicle \textit{v} in lane 2 in the scenario         \\ \hline
\multicolumn{1}{|l|}{\textit{slowSpeed(v)}}         & vehicle \textit{v} drops speed below the speed limit                  \\ \hline
\multicolumn{1}{|l|}{\textit{overSpeed(v)}}         & vehicle \textit{v} exceeds the speed limit in either lane             \\ \hline
\multicolumn{1}{|l|}{\textit{speed (v, $S_v$)}}        & momentary speed $S_v$ of vehicle \textit{v}                              \\ \hline
\multicolumn{1}{|l|}{\textit{acceleration (v, $A_v$)}} & momentary acceleration $A_v$ of vehicle \textit{v}                       \\ \hline
\multicolumn{1}{|l|}{\textit{deceleration(v, $D_v$)}}  & momentary acceleration $D_v$ of vehicle \textit{v}                       \\ \hline
\multicolumn{1}{|l|}{\textit{hardBraking (v)}}  & vehicle \textit{v} executes an emergency braking (hard braking deceleration (hbd) is -8 m/s2)   \\ \hline
\multicolumn{1}{|l|}{\textit{normalBraking(v)}} & vehicle \textit{v} executes normal braking (i.e., normal braking deceleration(nbd) is -3 m/s2 ) \\ \hline
\multicolumn{1}{|l|}{\textit{stopping (v)}}         & vehicle \textit{v} reduces its speed after braking                    \\ \hline
\multicolumn{1}{|l|}{\textit{safeDriving(v)}}       & vehicle \textit{v} is a safe driver                                   \\ \hline
\multicolumn{1}{|l|}{\textit{distractedDriving(v)}} & vehicle \textit{v} is a distracted driver                             \\ \hline
\multicolumn{1}{|l|}{\textit{aggressiveDriving(v)}} & vehicle \textit{v} is an aggressive driver                            \\ \hline
\end{tabular}
\vspace{-18pt}
\end{table}

Micro-behaviors such as acceleration, braking, lane change to the left or right, etc., are represented as events in EC that are defined along with their associated timestamps, which indicate the point in time at which the activity occurred. The \textit{hA} predicate establishes this type of input. For instance, \textit{hA(hardBreaking(id6, 60)} indicates that an object(id6) engaged in hard or sudden braking at video frame 60, which is determined by comparing the deceleration value going beyond a predefined threshold. Some of the micro behaviors are represented as fluents in the EC. We use the \textit{inA} and \textit{tA} predicates for expressing the conditions in which these fluents initiate and terminate a specific driver behavior described above.

The Micro-behaviors represented as EC events are defined mostly with \textit{hF} predicate, which can also compute the associated intervals. For example, \textit{hF(laneDrifting(id3) = true, [(0, 60),(210, 280 )]} indicates that object3 was not keeping the center of the lane during the intervals (0, 60) and (210, 280). A few examples of events and fluents defined along with their meaning are shown in  Table~\ref{tab:evt-fluents}. 

After defining events and fluents, we must define an initiation and termination map for each defined fluent in the system, indicating which events initiate and terminate which fluents. The next step is to specify the relation between fluents and events in the form of rules. For example, the initiation and termination map for fluent \textit{overSpeed(v)} is shown in Definition~\ref{f:overspeed}. In the definition, \textit{overSpeed}, \textit{speed} are input events and \textit{lane1} and \textit{lane2} are the input fluents. \textit{th} is an temporal predicate indicating numerical threshold of driving parameters and in this case it represents the user-specified speeding threshold. The ruleset defined states that \textit{overSpeed(v)} is a Boolean fluent, which is invoked when a \textit{speed} event is detected. Further, the vehicle is present in lane1 and not in lane2, which is detected by fluents \textit{lane1} and \textit{lane2} respectively, and the momentary speed of the vehicle is more than the user-specified speeding threshold. The event \textit{overSpeed(v)} is terminated when the vehicle vs. speed is smaller than the speeding threshold. The exact definition applies when the vehicle is located in \textit{lane2}, as indicated by fluent \textit{lane2} and the negation of fluent \textit{lane1}.

\begin{Definition}[ht]
\vspace{-5pt}
\small
\hrulefill

    \textit{inA(overSpeed(v) = true, T)  $\leftarrow$ \textit{hoA(speed(v, $S_{v}$), T)} $\land$
    {hoA(atLane1(v), T)} $\land$
    {$\neg$ hoA(atLane2(v), T)} $\land$ \\
    \textit{th($os$, $S_{v}$ $>=$ $os$)}}.


    \textit{tA(overSpeed(v) = true, T)}  $\leftarrow$ \textit{hoA(speed(v, $S_{v}$), T)} $\land$
    {\textit{hoA(lane1(v), T)}} $\land$
    {\textit{$\neg$ hoA(lane2(v), T)}} $\land$ \\ 
    \textit{\textit{th($os$, $S_{v}$ $<$ $os$)}}.

\hrulefill\vspace{-0.1in}
\caption{Initiation and Termination map of fluent \textit{overSpeed}}
\label{f:overspeed}
\vspace{-6pt}
\end{Definition}

\subsection{Characterizing Driver Behavior}

As discussed in~\ref{s:catergories-driver}, each category of driver behavior is a combination of different types of micro-behaviors, wherein all micro-behaviors defining a driver's behavior must be satisfied. For instance, the micro-behaviors that represent aggressive driver behavior as RTEC events or fluents include events such as suddenSteer, weaving, as well as fluents such as laneChange, atLane1, atLane2, overSpeed, and hardBraking. The aggressive driving behavior is represented as a boolean fluent defined as shown in the Definition~\ref{f:aggressive}. Similarly the distracted driver behavior is a boolean fluent defined as conjunction of events like laneDrifting, straddling, and fluents like laneChange, atLane1, atLane2, slowSpeed and normalBraking. 

\begin{Definition}[ht]
\vspace{-5pt}
\small
\hrulefill

    \textit{inA(aggressiveDriving(v) = true, T) $\leftarrow$ {hoA(atLane1(v), T)} $\land$ \\
    {$\neg$ hoA(atLane2(v), T)} $\land$ 
    {hoA(hardBraking(v), T)} $\land$ \\
    {hoA(laneChange(v), T)} $\land$ 
    {hoA(overSpeed(v), T)} $\land$ \\
    {hA(weaving(v), T)} $\land$ 
    {hA(suddenSteer(v), T)} $\land$ \\
    {$\neg$ hA(safeDriving(v), T)} $\land$  
    {$\neg$ hA(distractedDriving(v), T)} 
    }.

\vspace{0.05in}    
\hrulefill\vspace{-0.1in}
\caption{Characterization of \textit{Aggressive} Driver Behavior}
\vspace{-5pt}
\label{f:aggressive}
\end{Definition}

To express the dependencies when modeling the driver behavior, we define derived events, i.e., the events that occur due to the change in state or value of another fluent and/or the occurrence of another event (e.g., the effect of a normal or harsh braking on a vehicle's speed). These events indicate when specific actions occur in the traffic due to a combination of particular conditions. If $\textbf{R}$ denotes the rules passed to the WPM2 solver, before invoking the WPM2 solver, we find relevant rules corresponding to derived events based on current ongoing events~\cite{Pavana-IoT}. We identify the rules or relations $\textbf{R'} \subseteq \textbf{R}$ that lead to derived events based on the current events, either directly or indirectly. The dependency is expressed via a dependency graph $G$, where the vertices denote the rules/relations, and the (directed) edges denote the dependency between them. We then take the transitive closure of $G$ (say $G'$) using the Floyd–Warshall algorithm. Thus, in $G'$ an edge $i \rightarrow j$ denotes that $j$ is directly or indirectly dependent on $i$. Finally, the rules $\textbf{R'}$ expressed in CNF are then passed on to a WPM2 solver~\cite{ansotegui2010new}.

\section{Experimental Evaluation}
\label{s:experimental-evaluation}

In this section, we evaluate our framework on our collected PiCarX dataset. The effectiveness of our framework is determined by the following metrics: (a) the accuracy and (b) the percentage of errors in driver behavior recognition. The experiments were performed on a computer with Intel(R) Core(TM) i7-7700 CPU @ 3.60 GHz, 32 GB RAM, and 1 TB SSD and SWI-Prolog 8.2.3.

\subsection{Modeling Vehicular Driving}

The vehicular traffic on a roadway has been characterized by numerous models starting with the early 1950s. The models can be microscopic (i.e., model the behavior of each vehicle) or macroscopic (i.e., model the behavior of the traffic as a whole). Numerous microscopic (or ``car-following") models exist, which are reviewed in a recent article~\cite{saifuzzaman2014incorporating} that also examines the incorporation of human factors into these models. Most models are continuous time, continuous space type, and for a single lane, express acceleration of a vehicle as a function of its current speed, distance to and speed of next vehicle (and sometimes the previous vehicle as well), etc. Most models also introduce some random slowdowns to model human behavior and to break the strict vehicle following behavior. Generally, lane change behavior is tacked on to the single-lane models with a set of rules concerning when lane change can occur; however, such models quickly become very difficult to analyze mathematically.  

In our implementation, we used the so-called cellular automaton (CA) model, which simplifies the introduction of complex rules and simulation implementation by discretizing both time and space. In the CA model, a roadway is seen as a sequence of cells of some fixed size. At any point in time, a cell could be empty or occupied. That is, a cell can be occupied by only one vehicle, although it is possible to model large vehicles that occupy multiple consecutive cells. In each time step, a vehicle may move over some integral number of cells depending on its speed and the availability of the cells ahead. The CA model makes it easy to introduce complex rules that account for various situations, including the presence of signals or other traffic control mechanisms. Complex lane change rules can also be coded; for example, coding a lane change decision based on the number of free cells ahead and behind in both the source and target lanes. It is typical to assume that a lane change always occurs in a single step, and the discrete model is well suited for this kind of discontinuous change. CA model makes it easy to provide various forms of driving personalities to a vehicle in terms of vehicle following, lane change, safe/unsafe vehicle position in a cell, etc.

\subsection{Experimental Setup and Dataset Collection}

One significant difficulty in conducting this work was the scarcity of real datasets that provide both the vehicular motion parameters and the external videos that we can analyze. 

This paper can be considered a proof of concept (PoC) of characterizing the vehicular behavior remotely and non-intrusively without any involvement of the vehicle/driver or deployment of any further instrumentation in the vehicles. In particular, this paper aims to study how accurately this can be done by comparing it against the ground truth. Unfortunately, it is impossible to conduct such a study with actual vehicles on the road because of safety concerns since we need the vehicles to perform unsafe maneuvers. Also, to compare the video monitoring results against the ground truth, we need to tap into the OBD and obtain detailed vehicular information.

In order to get around these difficulties, we set up an entire infrastructure using the automated toy cars, known as ``PiCarXs"~\cite{PicarX} in an indoor basement environment. Each PiCarX carries a raspberry pi board for programmatically controlling the car. We assembled six such cars and used a logically centralized controller to independently control each car's behavior remotely over the WiFi link. We also set up a ``roadside" external camera to record the videos of the cars independently and analyze those videos in real time to determine and classify the behavior of each PiCarX. Each PiCarX used the 2-lane extension of the basic Cellular Automata (CA) car-following model~\cite{wolfram2003new} and further updated the model to introduce different types of driver behaviors. The software control of the vehicles can read their current speed and other parameters via appropriate sensors. We can then analyze the driver behavior for both in-vehicle sensing units and roadside camera units using a mixture of CNN models and logic-based behavior analysis. Finally, we validate our claim of analyzing driver behavior ``only'' from the roadside units by correlating the in-vehicle data and the roadside data. 

Thus, we collected such a dataset using our PiCarX setup. We captured High Definition Video Streams (HDVS) at 30 FPS with PicarX cars imitating different driver behavioral patterns on the road as discussed in section~\ref{s:catergories-driver}. By varying speed limits, we also capture the driving traffic patterns on different road types, such as highways versus local roads. Along with capturing the video streams, we also record the eight in-vehicular driving parameters from all six robot cars, including vehicle orientation, acceleration, deceleration, braking, steering angle, lateral position, and lane change maneuvers to the left and right lanes.

\begin{figure}[htb]
\center
\subfigure[Front View of PiCarX]{\includegraphics[width=1.55in,angle=0]{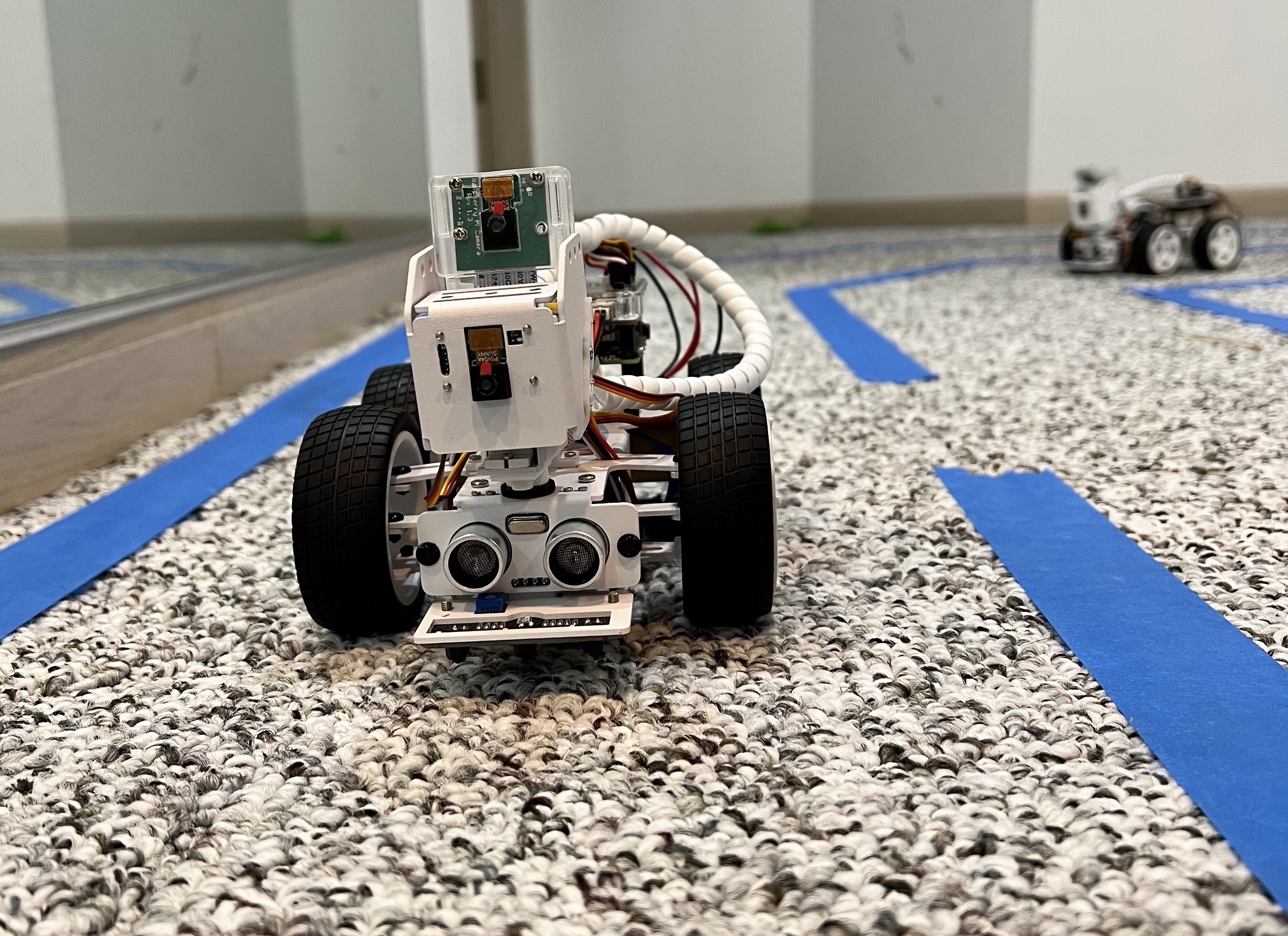}}~
\subfigure[Side View of PiCarX]{\includegraphics[width=1.50in,angle=0]{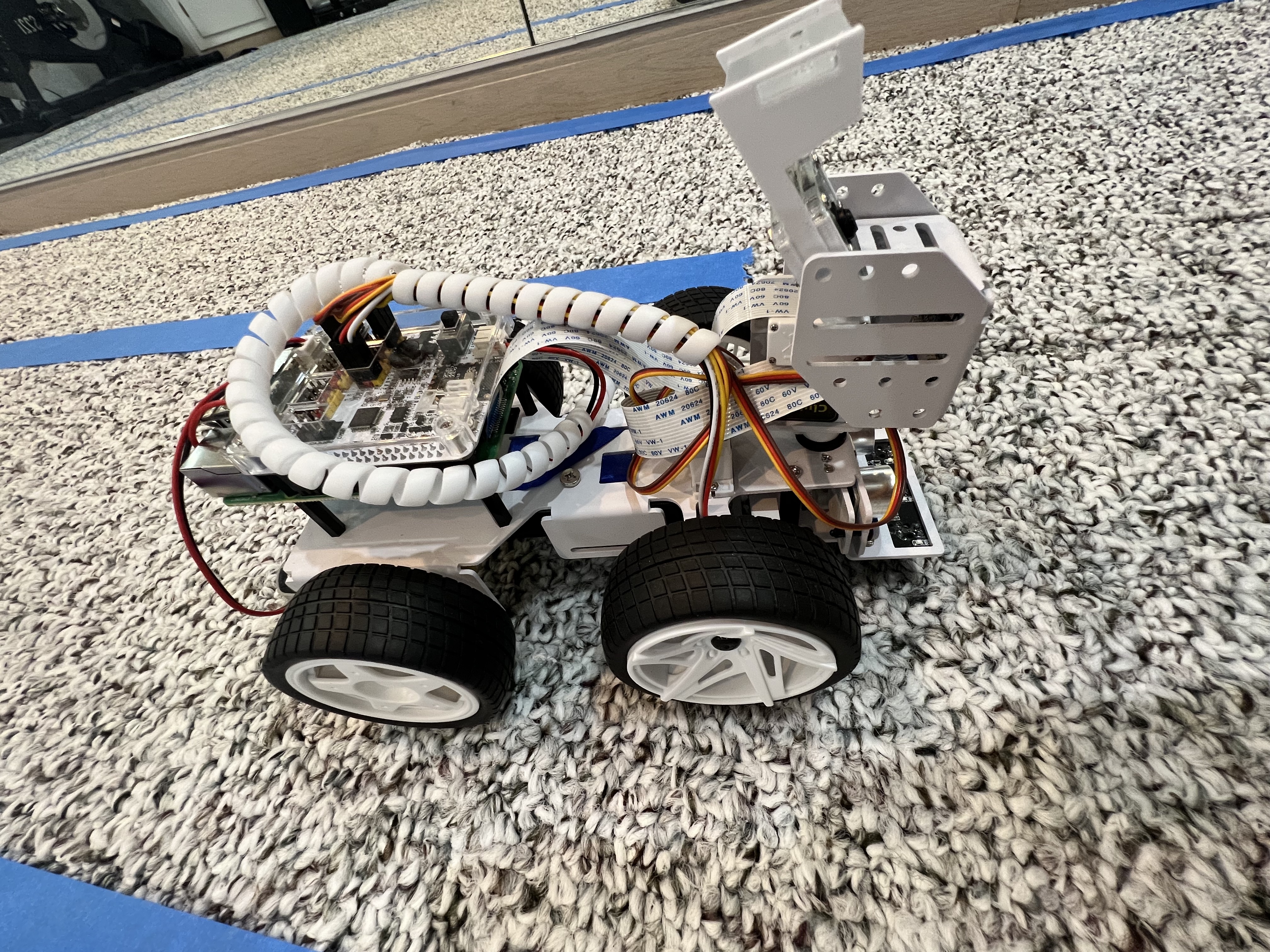}}
\vspace{-3pt}
\caption{PiCarX used in the experiments.}
\vspace{-15pt}
\label{f:picarx}
\end{figure}

Fig.~\ref{f:picarx}(a) and (b) shows the front and side view of PicarX used in the experiments, respectively. The experimental setup is shown in Fig.~\ref{f:picarx-detection} and shows two distinct lanes with a single direction flow of traffic. We use a tripod and Sony FDR-AX33 Camcorder to record the simulation videos.

We conducted five video simulations for our controlled environment, with each video being around 1-2 minutes long and averaging around 10,000 video frames. In addition, to simulate the in-vehicle framework as discussed in section~\ref{s:overall-framework}, we have approximately 2700 seconds of time-series data for all required in-vehicle driving parameters as discussed above. 

In addition to the simulated videos, we select approximately 48 real-world traffic videos from our other dataset, TU\_DAT dataset to evaluate the effectiveness of our proposed work.
TU\_DAT dataset was used in our previous work~\cite{Pavana-CFAR} to predict and resolve anomalous situations in CCTV traffic-videos, which contain a diverse collection of accident videos collected in challenging environments. Table~\ref{tab:char-datasets} shows the details of the data collected and the number of samples. 

\subsection{Implementation of Cellular Automata (CA) Model}

We implement the CA car-following model in each PiCarX and then extend the model to implement different driving behaviours.
A classical CA model is a uniform lattice of cells representing an identical finite automaton with a state and a transition function; each cell takes its state and the states of a set of neighboring cells defined by a time and space invariant geometrical pattern. Starting with an initial condition, the CA evolves by sequentially activating all transition functions simultaneously. We use a generic, multi-layered and complete cellular automaton simulation engine in Python called ``Cellular Automata General Environment” (CAGE)~\cite{blecic2004generalized}. The environment supports multilayered grids, which provide several ways for the formation of neighborhoods. It enables the definition of transition rules that can be formulated algorithmically to reflect real-time driver behavior. Furthermore, the rules can be specified at different spatial levels and can change as a function of space and time, making it suitable for use alongside a logical reasoning tool. 

In our two-lane experimental setup, we implement abstractions of the CA model, road topology, and neighborhood information using CAGE to represent the various driver behaviors and a two-lane car-following model. An address in CAGE is a tuple of one or more integers representing the location of a cell within a given topology. Topologies determine the arrangements of cells in a network, and neighborhoods encapsulate the translation of addresses to a list of their neighbors, taking into account the topology they are connected with. The Map is the high-level class used by the automaton to perform operations on the cellular network. It is a combination of the topology and the Neighborhood classes. For the purposes of our experiments, we implemented a line map topology with a radial neighborhood for each lane having a total of 35 cells.

\begin{table}[]
\center
\vspace{-18pt}
\caption{Statistics of Datasets Used}
\label{tab:char-datasets}
\begin{tabular}{|c|l|l|}
\hline
\multicolumn{1}{|l|}{Data Collected} & Behavior & \#Samples \\ \hline
\multirow{3}{*}{\begin{tabular}[c]{@{}c@{}}Road-Side Camera \\ Video Data (PicarX) \end{tabular}} & Safe & 410 \\ \cline{2-3} 
 & Distracted & 340 \\ \cline{2-3} 
 & Aggressive & 300 \\ \hline 
\multirow{3}{*}{\begin{tabular}[c]{@{}c@{}}In-Vehicular \\ Data (PicarX) \end{tabular}}
& Safe & 125 \\ \cline{2-3} 
 & Distracted & 115 \\ \cline{2-3} 
 & Aggressive & 110 \\ \hline
\multirow{3}{*}{TU-DAT Dataset} & Safe & 1200 \\ \cline{2-3} 
 & Distracted & 720 \\ \cline{2-3} 
 & Aggressive & 960 \\ \hline
\end{tabular}
\vspace{-18pt}
\end{table}
\subsection{PicarX Object Detection}

\begin{figure}[t]
\center
\subfigure[Without object detection]{\includegraphics[width=1.55in,angle=0]{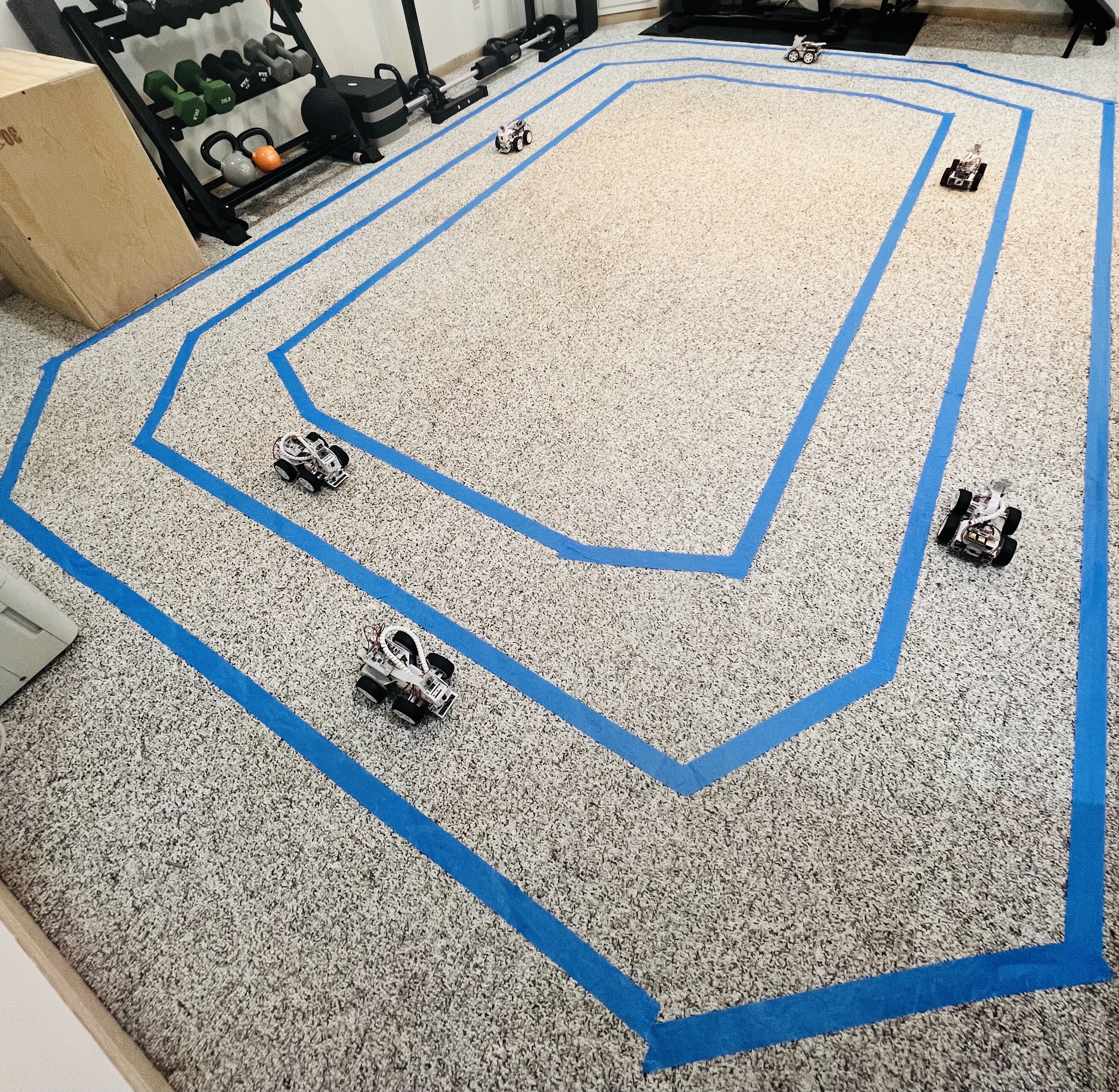}}~
\subfigure[PicarX detection and tracking]{\includegraphics[width=1.55in,angle=0]{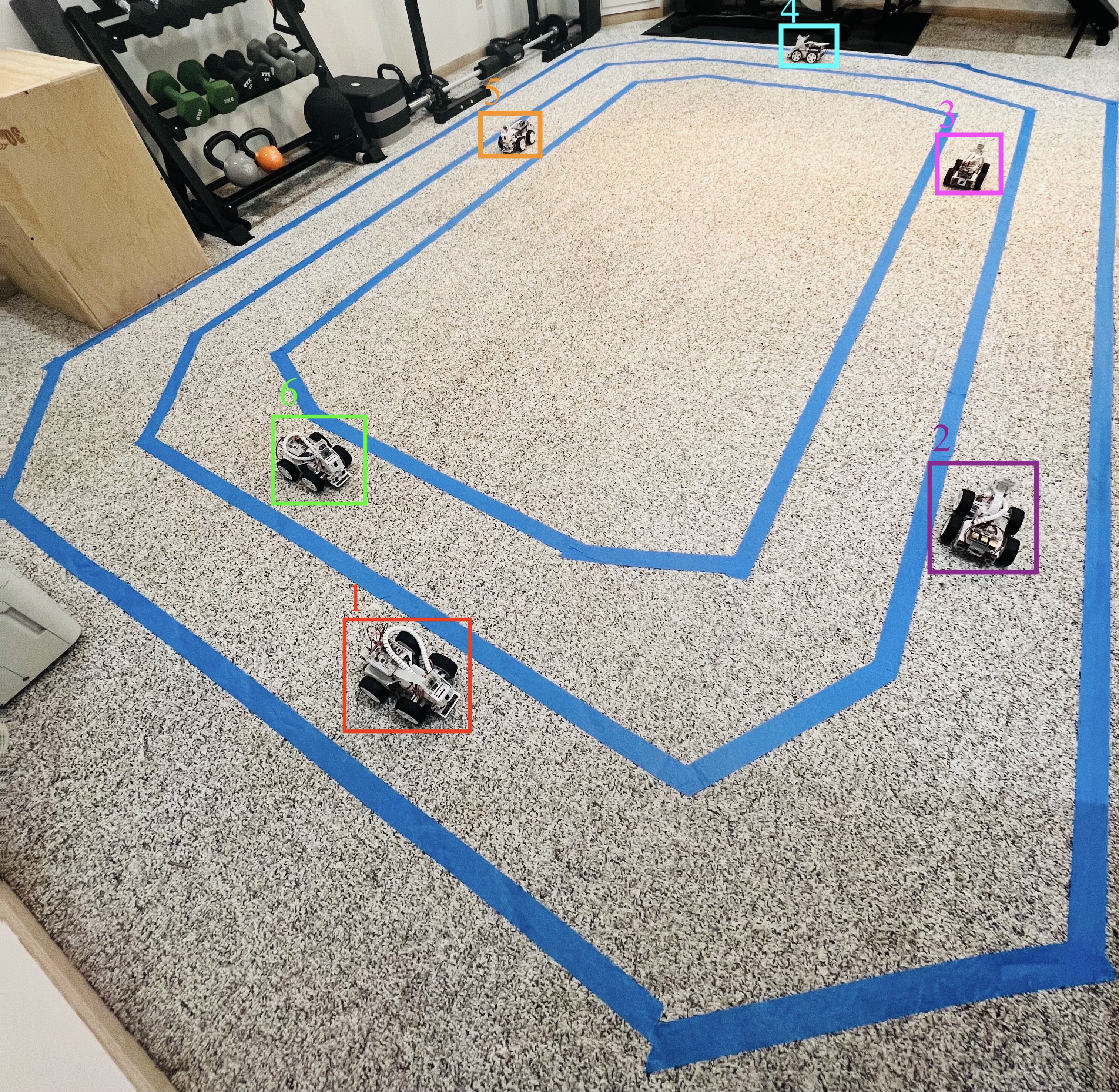}}
\vspace{-3pt}
\caption{PiCarX object detection and tracking.}
\vspace{-18pt}
\label{f:picarx-detection}
\end{figure}

As discussed in~\ref{s:yllo}, the roadside camera or RECs use YLLO-based object detection and tracking framework and a logical-reasoning-based method to classify the driver's behavior into one of three categories. The YLLO running on the cameras must be able to detect and classify the robot cars since we are using robot cars to simulate road traffic consisting of different driver behaviors. To train YLLO to recognize the robot cars, we constructed our own training and testing set by selecting positive samples from the recorded videos. For negative samples, we have utilized the frames from our toy car experiment videos used in our previous work~\cite{Pavana-YLLO}. Overall we have a total of 18,550 samples used to train the model. To reduce the workload of annotating the dataset, we have used an auto annotation tool~\cite{Auto}, which is based on a semi-supervised architecture, where a model trained with a small amount of labeled data is used to produce the new labels for the rest of the dataset. Since YLLO is a CNN-based object detection model, it requires large amounts of data. Hence, we have used various augmentation techniques to reach sufficient data amounts, with most augmentations occurring at run-time using Keras built-in functions. Keras offers several techniques for performing image augmentation online, which means that the augmentation is done as each image is processed by the network. Consequently, multiple augmentations can be performed without the need to save each image separately on the computer. The augmentations used were flipping, translation, shear, and rotation. The trained model has an overall identification accuracy of 90.11\%. Fig.~\ref{f:picarx-detection} shows the PiCarX detection and tracking results of the trained YLLO model. Due to space limitations, the trained model's results are omitted from this paper. 

We evaluate the performance of the proposed driver behavior detection system using both roadside cameras and an in-vehicle framework. We evaluate the classification performance of the driver behavior using standard performance metrics such as precision, recall, and accuracy. 

\subsection{Results and Discussion}

Estimating various vehicular parameters, such as speed, vehicle orientation, etc., forms the basis of driver behavior analysis~\cite{hu2004survey}. For this, we need to first calibrate the camera so that it is possible to correct the inherent perspective distortion in the images. The perspective effect relates 3D points on the road (world) coordinate system to 2D pixels on the image plane differently. This effect assigns distinct informational contents to different image pixels. The objective of Inverse Perspective Mapping (IPM) is to invert the perspective effect, thereby imposing a uniform distribution of information across the image plane. To map the front-view image smoothly into a bird's-eye view for videos captured with PicarX cars, we employ the IPM technique.

\begin{table}[htb]
\vspace{-8pt}
\caption{Performance Results using PicarX Dataset}
\label{tab:picarx-results}
\begin{tabular}{|p{.65in}|p{.45in}|p{.55in}|p{.4in}|p{.55in}|}
\hline
\multicolumn{1}{|l|}{Data Collected} & Driving Behavior & Precision & Recall & Accuracy \\ \hline
\multirow{3}{*}{\begin{tabular}[c]{@{}c@{}}Road-Side \\ Camera  \end{tabular}} & Safe & 100 & 98.78 & 99.52 \\ \cline{2-5} 
 & Distracted & 95.95 & 98.22 & 98.09 \\ \cline{2-5} 
 & Aggressive & 97.98 & 97.0 & 98.57 \\ \hline
\multirow{3}{*}{\begin{tabular}[c]{@{}c@{}}In-Vehicular \\ Data\end{tabular}} & Safe & 100 & 99.2 & 99.71 \\ \cline{2-5} 
 & Distracted & 98.28 & 99.13 & 99.14 \\ \cline{2-5} 
 & Aggressive & 99.09 & 99.09 & 99.43 \\ \hline
\end{tabular}
\end{table}

The performance of the proposed roadside camera-based driver behavior analysis framework is evaluated by first detecting a set of micro behaviors, which are then composed to classify the driver behavior into safe, distracted, and aggressive driving behaviors. We validate the proposed framework against the logic-based reasoning framework operating in individual vehicles, which provides the actual values of driving parameters such as speed, orientation, distance from the car in front, distance from the left or right side of the lane, etc. Table~\ref{tab:picarx-results} shows the results, and it can be seen that the roadside camera framework achieves an average precision and recall of 97.98\% and  98,005\%, respectively, and an average accuracy of 98.73\% which are averaged over five runs of the experiments. The difference between these performance metrics values and those of the in-vehicle framework is quite small, ranging between 1-2 percent.

In addition, we analyzed the errors that may arise when characterizing driver behavior via roadside cameras. 
\begin{wrapfigure}[10]{r}{0.60\columnwidth}
\vspace{-6pt}
\includegraphics[width=\linewidth]{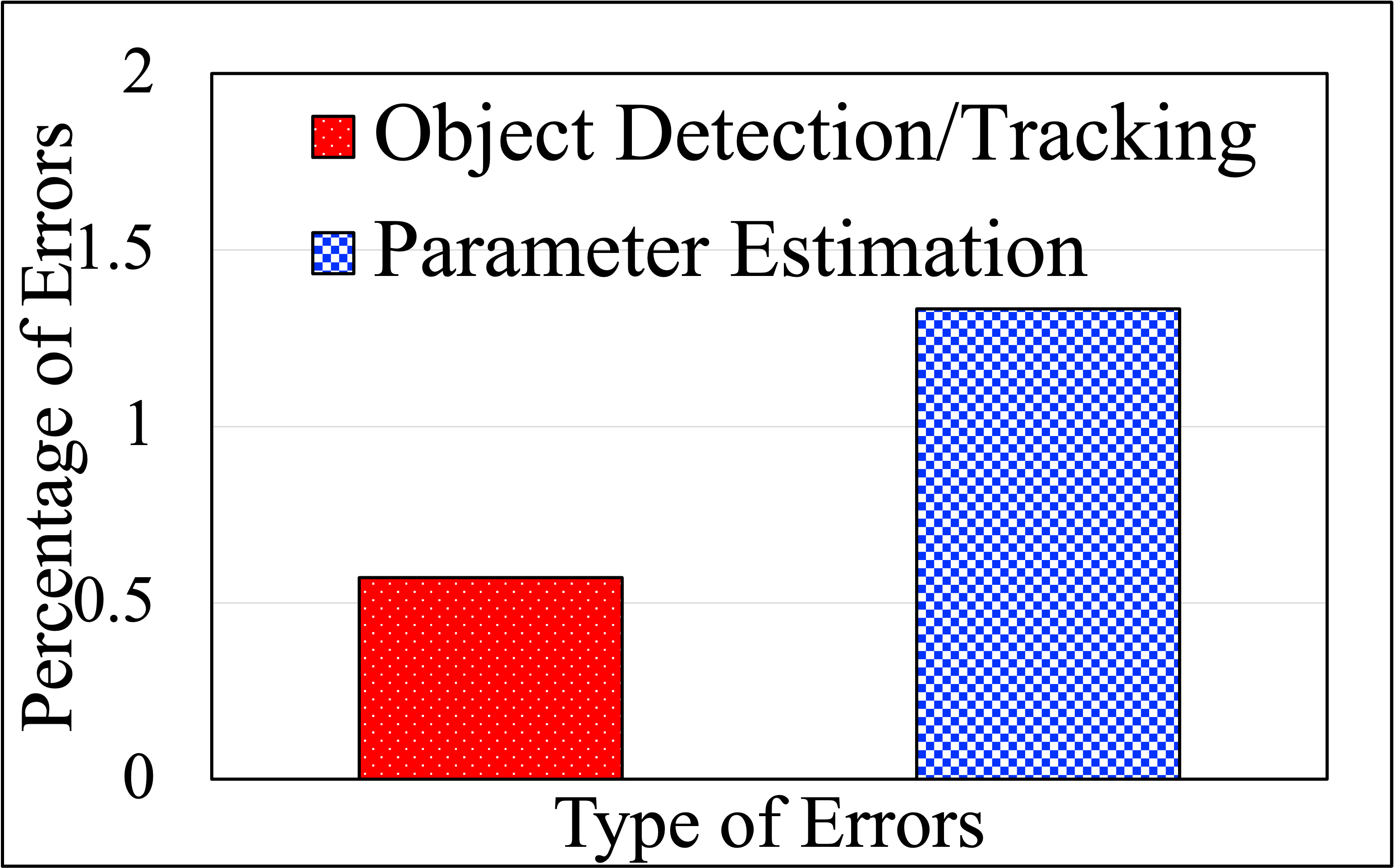}
\vspace{-20pt}
\caption{Error analysis.} 
\label{f:error}
\end{wrapfigure}
The outcome of this analysis is depicted in Fig.~\ref{f:error}, where the x-axis represents the types of errors and the y-axis represents the percentage of errors across all runs of the experiments described above. It can be seen that object detection and tracking errors are approximately 0.57\%, and driving signal parameter estimation errors in comparison to the ground truth values are approximately 1.33\%. Errors are nearly negligible, demonstrating the effectiveness of roadside camera-based driver behavior characterization. Based on our analysis, we conclude that the proposed framework can be used effectively to determine a driver's non-intrusive behavior without requiring the installation of additional sensors within the vehicles.

\begin{table}[htb]
\vspace{-12pt}
\caption{Performance Results using TU\_DAT (without IPM)}
\label{tab:tudat-results}
\begin{tabular}{|c|l|l|l|l|}
\hline
\multicolumn{1}{|l|}{Dataset} & Driving Styles & PRECISION & RECALL & ACCURACY \\ \hline
\multirow{3}{*}{TU-DAT} & Safe & 95.0 & 95.0 & 95.83 \\ \cline{2-5} 
 & Distracted & 83.5 & 84 & 91.7 \\ \cline{2-5} 
 & Aggressive & 93.75 & 93.75 & 95.8 \\ \hline
\end{tabular}
\end{table}

In addition to the recorded videos using PicarX, we have evaluated the proposed mechanism using the TU\_DAT dataset. Since this dataset is comprised of CCTV traffic videos of anomalous situations from various parts of the world, these cameras' intrinsic and extrinsic parameters may be unknown or different from one another due to mounting setups and camera types. Therefore, individual calibration of the captured videos is not possible. Table~\ref{tab:tudat-results} shows the performance results of the proposed roadside camera framework on TU\_DAT, it can be seen that the average precision and recall values are 90.694\%, and the accuracy is approximately 94.5\%. {\em It is important to note that the worse accuracy here is primarily due to lack of perspective correction; if the camera position and angles were known, we believe that the accuracies here would be similar to those obtained using PiCarX.}

\begin{table}[htb]
\vspace{-18pt}
\caption{Contribution of Micro-behavior to Aggressive/Distracted Driving after and before driver feedback\label{t:feedback}}
\begin{tabular}{|l|l|l|}
\hline
\textbf{Micro Behaviors} & \textbf{Aggressive} & \textbf{Distracted} \\ \hline
Close to  left side of the lane & 0.74 & 0.73 \\ \hline
Close to right side of the   lane & 0.77 & 0.73 \\ \hline
Drifting from center of the   lane & 0.7 & 0.59 \\ \hline
Too near to a vehicle in the   same lane & 0.71 & 0.69 \\ \hline
Too near to a vehicle in the   adjacent lane & 0.7 & 0.84 \\ \hline
Speeding & 0.61 & N/A \\ \hline
Harsh braking & 0.69 & N/A \\ \hline
Changing Lanes & 0.64 & 0.62 \\ \hline
Sudden steering & 0.66 & N/A \\ \hline
Slow Speed & N/A & 0.62 \\ \hline
\end{tabular}
\vspace{-20pt}
\end{table}

Table.~\ref{t:feedback} shows how the change in micro-behavior by a driver as a result of feedback can reduce the instances of aggressive and distracted driving. Each row in this table represents the situation where the driver is informed about its undesired micro-behavior, and as result, the driver reduces that behavior by 50\%. (This is done one at a time for each row, not cumulatively). The columns show the aggressive and distracted driving occurrences that include the said micro-behavior. The reported quantity is the ratio of the instances after and before the feedback and thus shows the impact of the feedback. The Not Applicable (N/A) reference indicates that the corresponding micro-behavior is not considered when classifying a driver as aggressive or distracted. The table demonstrates that the most effective micro-behavior improvements for aggressive and distracted drivers are speed control and maintaining the center of the lane. 
Thus, such an analysis can tell us the most effective feedback for manual or automated vehicles.

\section{Conclusions and Discussion}
\label{s:conclusions}

The ITS support involves, at a minimum, roadside traffic monitoring infrastructure with monitoring nodes deployed on the electric poles and usually has cameras and some computing and backhaul communications capability to send raw or processed video data to a central entity such as a cloud. Such a minimal setup does not require any communications capability in the vehicles themselves, and we demonstrate in this paper, through emulation of such an environment, that such a capability is adequate to monitor and classify behaviors of various drivers in the camera view without any involvement of or awareness by either the car or the driver. As such, this behavior characterization can only be exploited for improving the signage on the road that warns the drivers or is used the change the requirements (e.g., speed limit, lane following requirements, traffic control options, etc.). 

Our proof of concept using automated robot cars demonstrates that driver behavior characterization can be done very accurately by only using the view from the roadside cameras. We believe this is a significant result and paves the way for further development of the technique. Two notable additions would be tracking vehicles through a rich feature set and fusing views from multiple cameras to cover areas that cannot be covered well by a single camera. Another piece is the use of characterization in providing appropriate feedback to drivers and automated vehicles, although the form of feedback would be very different and needs to be studied. This obviously requires V2I (vehicle to infrastructure) communications capability, which is being implemented already. Automated vehicles, when fed with driving behavior information about nearby vehicles, can use it to reduce the traffic waves and other inefficiencies due to manually driven vehicles~\cite{SeiboldFlynnKasimovRosales2013}. The analysis can also be extended for more general environments, such as those with more than two lanes, bi-directional traffic, traffic signals, sharp curves, poor road conditions, etc.


\vspace*{-5pt}
\bibliographystyle{IEEEtran}
\bibliography{main.bib}

\end{document}